\documentstyle[12pt,epsf,epsfig,a4]{article}
\begin{document}
\begin {titlepage}
\begin{flushleft}
FSUJ TPI QO-08/96
\end{flushleft}
\begin{flushright}
June, 1996
\end{flushright}
\vspace{20mm}
\begin{center}
{\Large \bf Direct sampling of the Susskind--Glogower \\[.5ex]
phase distributions } 
\vspace{15mm}
 
{\large \bf M. Dakna, L. Kn\"oll, and D.--G. Welsch}\\[1.ex]
{\large Friedrich-Schiller-Universit\"at Jena}\\[0.5 ex]
{\large Theoretisch-Physikalisches Institut}\\[0.5 ex]
{\large Max-Wien Platz 1, D-07743 Jena, Germany}
\vspace{25mm}
\end{center}
\begin{center}
\bf{Abstract}
\end{center}
Coarse-grained phase distributions are introduced that approximate
to the Susskind--Glogower cosine and 
sine phase distributions. The integral relations between the phase 
distributions and the phase-parametrized field-strength distributions
observable in balanced homodyning are derived and the integral kernels
are analyzed. It is shown that the phase distributions can be directly
sampled from the field-strength distributions which offers the
possibility of measuring the Susskind--Glogower cosine and sine phase 
distributions with sufficiently well accuracy. Numerical simulations are
performed to demonstrate the applicability of the method.
\end{titlepage}
\renewcommand {\thepage} {\arabic{page}}
\setcounter {page} {2}
\section{Introduction}
\label{sec1}
Although the problem of defining and measuring quantum mechanical phases
of radiation-field modes has been discussed for a long time, there has 
been no unified approach to the phase problem so far \cite{Lynch1}.
The reason is that there is no ``good'' (i.e., self-adjoint) phase
operator in the Hilbert space. The difficulties obviously arise  
from the boundedness of the eigenvalue spectrum of the photon-number
operator, which is the canonically conjugate of the phase operator.

One way to overcome the difficulties was proposed by Susskind and
Glogower \cite{Susskind1} who used the non-Hermitian exponential
phase operator and its adjoint in order to define two Hermitian 
operators. From the analogy between them and classical trigonometric 
functions the two operators are also referred to as cosine and sine 
phase operators. In particular in the classical limit they exactly 
correspond to the cosine and the sine of the phase. Since the cosine 
and sine phase operators are self-adjoint, their eigenstates can 
be used to define proper probability distributions for observing
the cosine and sine phases. However, these distributions cannot
be measured, in general, simultaneously and do therefore not
uniquely characterize the phase of the quantum state of a mode. 
This is obviously the price to pay for introducing self-adjoint 
operators and keeping the concept of quantum mechanical probabilities.

In the classical limit the difference between the cosine and sine phase 
distributions vanishes. In particular, in classical optics 
the phase difference between two radiation-field modes can always be 
determined by simultaneously measuring the cosine and sine of the phase 
difference in interference experiments. Inspired by 
such kinds of measurements, Noh, Foug\`{e}res, and Mandel \cite{Noh1} 
introduced an operational approach to the cosine and sine phase operators. As 
expected, for classical fields the corresponding phase distributions agree 
with the quantum mechanical Susskind--Glogower (SG) phase distributions, 
but for quantum fields they may be quite different from each other.

A powerful interferometric method for determining phase-sensitive
properties of optical fields has been balanced homodyne detection. It 
is well known that from the data recorded in a succession of measurements
the quantum state of a signal mode can be obtained, as was first
demonstrated experimentally by Smithey, Beck, Faridani, and Raymer 
using optical homodyne tomography \cite{Smithey1}. Knowing the
quantum state (in terms of, e.g., the density matrix in a chosen basis),
the quantum phase statistics can be calculated \cite{Beck1}. 
In the present paper we show that the SG cosine and sine phase distributions 
can be measured more directly by sampling them from the recorded data
\cite{Dakna1}.

The approach is based on the introduction of parametrized cosine
and sine phase distributions that are defined on the basis of 
appropriately coarse-grained SG cosine and sine quantum states. 
So, the parametrized distributions can be regarded as
smoothed SG cosine and sine phase distributions that
reflect the exact statistics in a coarse-graining approximation.
The accuracy is determined by the parameter that defines
the phase coarse-graining interval. In particular, the
parametrized distributions tend to the exact ones
when the coarse-graining parameter approaches zero.

The advantage of the method is that for any non-zero value of the
parameter the coarse-grained sine and cosine phase distributions
can be directly sampled from the recorded data in balanced homodyning. Hence, 
choosing the coarse-graining interval small enough, the exact distributions can 
be measured with sufficiently well accuracy. Since in balanced homodyning 
the observed difference-count statistics is (for chosen phase parameters of 
the apparatus) a scaled field strength distribution of the signal mode, the 
sampling function needed is given by the kernel in the integral relation 
between the parametrized cosine and sine phase distributions and the 
phase-parametrized field-strength distributions.
     
The paper is organized as follows. In Sec.~\ref{sec2} coarse-grained cosine 
and sine phase distributions are introduced. Section \ref{sec3} is 
devoted to the relations of the phase distributions to the field-strength
distributions and the calculation of the cosine and sine
phase sampling functions. In the Sec.~\ref{sec4} results of
computer simulations of measurements of the cosine and sine
phase distributions for a squeezed vacuum state are presented,
and a summary and some concluding remarks are given in Sec.~\ref{sec5}.

\section{Coarse-grained cosine and sine phase distributions}
\label{sec2}
Using the exponential phase operator
\begin{equation}
\hat{E} = \sum_{n=0}^{\infty}|n\rangle\langle n+1|
\label{1.0}
\end{equation}
($\hat n$ $\! |n\rangle$ $\!=$ $\!n$ $\!|n\rangle$, $\hat n$ $\!=$
$\!\hat a^\dagger \hat a$, $[\hat a,\hat a^\dagger]$ $\!=$ $\!1$),
Susskind and Glogower \cite{Susskind1} introduced the Hermitian
cosine and sine operators 
\begin{equation}
\hat{C}
=\textstyle\frac{1}{2}(\hat{E}+\hat{E}^\dagger)
\label{1.1}
\end{equation}
and
\begin{equation}
\hat{S}
= - \textstyle\frac{1}{2} i (\hat{E}-\hat{E}^\dagger),
\label{1.1a}
\end{equation}
respectively, which satisfy the eigenvalue equations 
\begin{equation}
\hat{C}\,|\cos\phi\rangle=\cos\phi\,|\cos\phi\rangle
\label{1.2}
\end{equation}
($0$ $\!\leq$ $\!\phi$ $\!\leq$ $\!\pi$) and
\begin{equation}
\hat{S}\,|\sin\phi\rangle=\sin\phi\,|\sin\phi\rangle
\label{1.2a}
\end{equation}
($-\pi/2$ $\!\leq$ $\!\phi$ $\!\leq$ $\!\pi/2$).
The SG cosine and 
sine phase states $|\cos\phi\rangle$ and $|\sin\phi\rangle$, respectively, 
form orthonormal and complete sets in the Hilbert space, that is to say,
\begin{equation}
\langle\cos\phi|\cos\phi'\rangle=\delta(\phi-\phi'),
\label{1.3}
\end{equation}
\begin{equation}
\int_{0}^{\pi}\;{\rm d}\phi\;|\cos\phi\rangle\langle\cos\phi|=\hat{I}
\label{1.3a}
\end{equation}
and
\begin{equation}
\langle\sin\phi|\sin\phi'\rangle=\delta(\phi-\phi'),
\label{1.4}
\end{equation}
\begin{equation}
\int_{-\pi/2}^{\pi/2}\;{\rm d}\phi\;|\sin\phi\rangle\langle\sin\phi|=\hat{I}.
\label{1.4a}
\end{equation}
In order to give a unified approach to the states, let us
consider the $\psi$-parametrized Hermitian operator 
\begin{equation}
\hat{C}(\psi)
      =\textstyle\frac{1}{2}(\hat{E}e^{-i\psi}+{\hat{E}^\dagger}e^{i\psi}).
\label{1.4b}
\end{equation}
Recalling Eq.~(\ref{1.0}), $\hat{C}(\psi)$ is easily proved to satisfy the
eigenvalue equation
\begin{equation}
      \hat{C}(\psi)\,|\Phi,\psi\rangle = \cos\Phi\,|\Phi,\psi\rangle,
\label{1.4c}
\end{equation}
where 
\begin{equation}
      |\Phi,\psi\rangle = \sqrt{\frac{2}{\pi}}
      \sum_n e^{in\psi}\sin[(n\!+\!1)\Phi]\,|n\rangle.
\label{1.4d}
\end{equation}
Note that $|-\Phi,\psi\rangle$ $\! = $ $\!- |\Phi,\psi\rangle$ and
$|\Phi\!+\!\pi,\psi\rangle$ $\! =$ $\!|\Phi,\psi\!+\!\pi\rangle$. 
For chosen $\psi$ the states $|\Phi,\psi\rangle$, 
$0$ $\!\leq$ $\!\Phi$ $\!\leq$ $\!\pi$, form an orthonormal complete basis. 
The SG cosine and sine phase states in Eqs.~(\ref{1.2}) and (\ref{1.2a}), 
respectively, can be obtained by appropriately specifying the
states $|\Phi,\psi\rangle$,  
\begin{equation}
      |\cos\phi\rangle = \, |\Phi\!=\!\phi,\psi\!=\!0\rangle,
\label{1.4e}
\end{equation}
\begin{equation}
      |\sin\phi\rangle = \,
\left|\Phi\!=\!\textstyle\frac{1}{2}\pi\!-\!\phi,
\psi\!=\!\textstyle\frac{1}{2}\pi\right\rangle.
\label{1.4f}
\end{equation}

With regard to measurements, we now introduce coarse-grained
states as
\begin{equation}
      |\Phi,\psi,\epsilon\rangle = \frac{1}{\sqrt{\epsilon}}
      \int\limits_{\Phi-\epsilon/2}^{\Phi+\epsilon/2} {\rm d}\Phi' \,
      |\Phi',\psi\rangle.
\label{1.5}
\end{equation}
Using Eq.~(\ref{1.4d}), the states 
$|\Phi,\psi,\epsilon\rangle$ can be given by
\begin{equation}
|\Phi,\psi,\epsilon\rangle 
= \sqrt{\frac{2\epsilon}{\pi}}
      \sum_n e^{in\psi} \sin[(n\!+\!1)\Phi]
      \, {\rm sinc}[(n\!+\!1)\epsilon/2]\,|n\rangle
\label{1.5a}
\end{equation}
(${\rm sinc}\,x$ $\!=$ $\!\sin x/x$).
They are normalized to unity, 
\begin{equation}
\langle\Phi,\psi,\epsilon|\Phi,\psi,\epsilon\rangle=1,
\label{1.8}
\end{equation}
and tend to the exact states $|\Phi,\psi\rangle$ as 
$\epsilon$ approaches zero,
\begin{equation}
\lim_{\epsilon\to 0}\frac{1}{\sqrt{\epsilon}}
|\Phi,\psi,\epsilon\rangle=|\Phi,\psi\rangle.
\label{1.9}
\end{equation}
The coarse-grained cosine and sine phase states, respectively,
are obtained as $|\cos\phi,\epsilon\rangle$ $\!=$
$|\Phi$ $\!=$ $\!\phi$, $\!\psi$ $\!=$ $\!0$, $\!\epsilon\rangle$ and
$|\sin\phi,\epsilon\rangle$ $\!=$
$|\Phi$ $\!=$ $\!\frac{1}{2}\pi$ $\!-$ $\!\phi$, $\!\psi$ $\!=$ 
$\!\frac{1}{2}\pi$, $\!\epsilon\rangle$ [cf. Eqs.~(\ref{1.4e}), (\ref{1.4f}), 
and (\ref{1.5})]. According to Eq.~(\ref{1.9}), they approach the exact SG
cosine and sine phase states in the limit when $\epsilon$ $\!\to$ $\!0$.

\begin{figure}
\centering\epsfig{figure=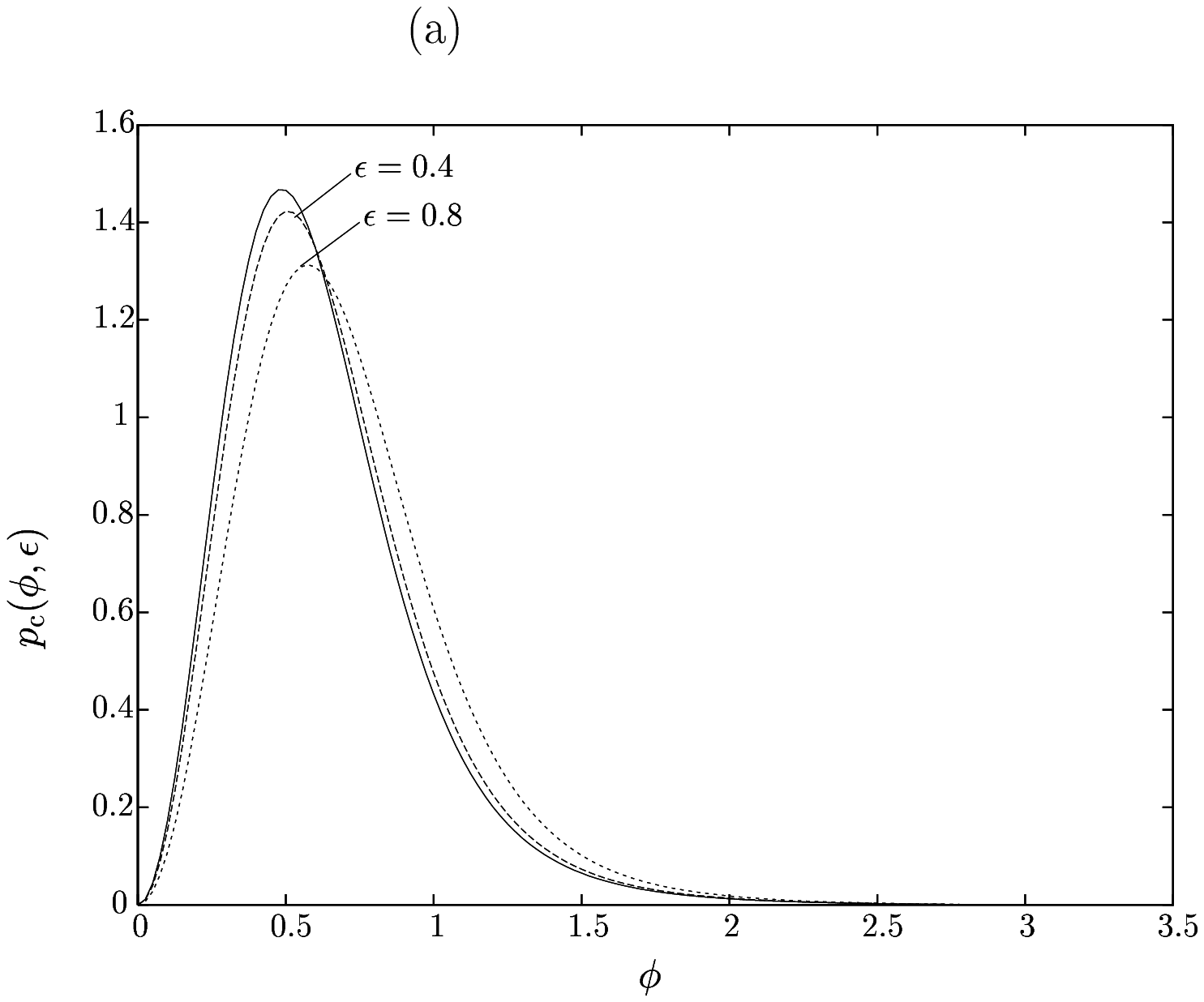,width=0.8\linewidth}
\centering\epsfig{figure=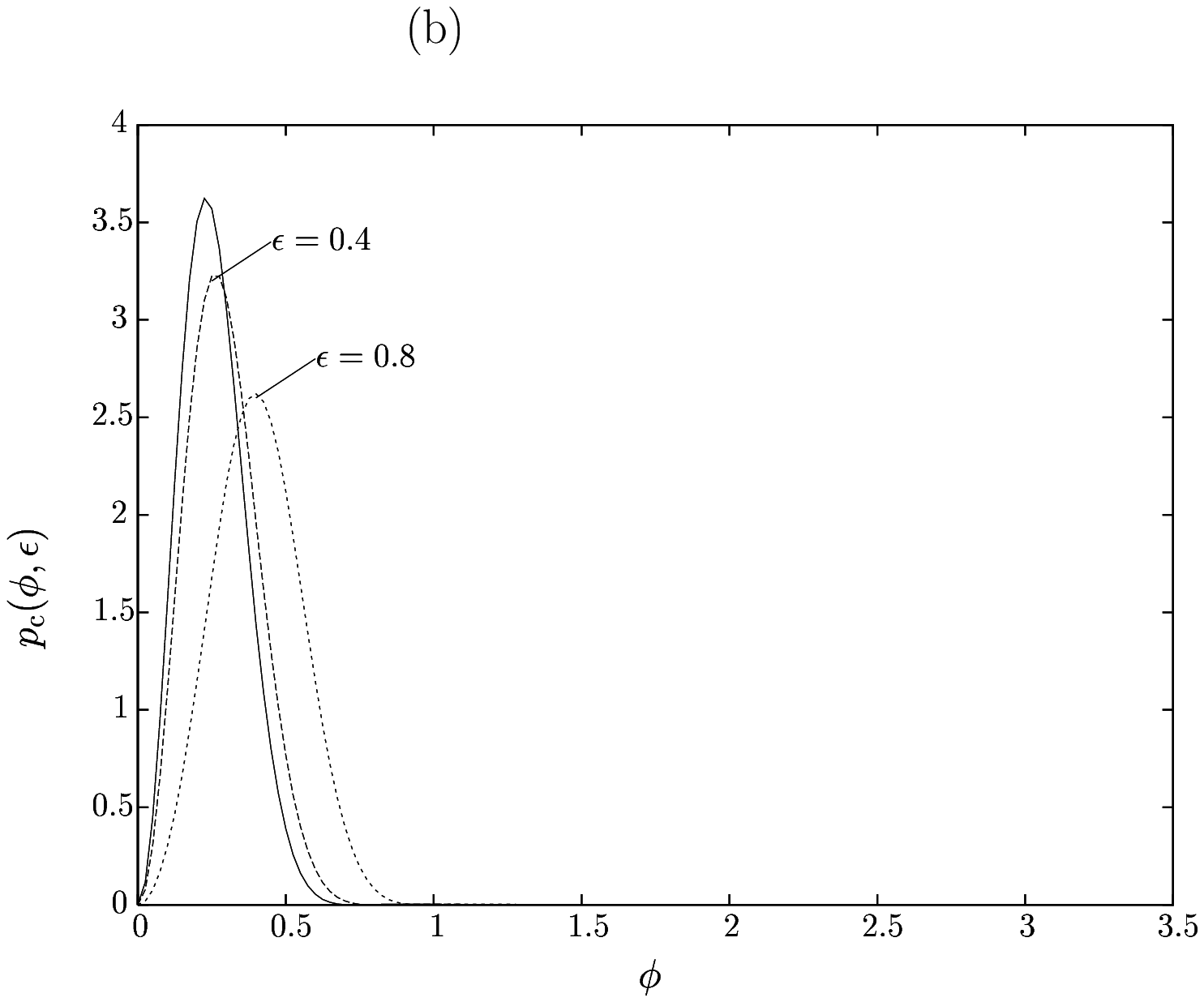,width=0.8\linewidth}
%\centering\epsfig{figure=pr13fig1.ps,width=0.8\linewidth}
%\centering\epsfig{figure=pr13fig2.ps,width=0.8\linewidth}
\caption{ \label{figr1}The coarse-grained cosine phase state
 distributions of
a mode prepared in a coherent state $|\alpha\rangle$ with
$\langle\hat n\rangle$ $\!=$ $\!1$ $(\alpha$ $\!=$ $\!1)$ (a) and 
$\langle\hat n\rangle$ $\!=$ $\!2$ $(\alpha$ $\!=$ $\!\protect\sqrt{2})$
(b) are shown for $\epsilon$ $\!=$ $\!0.4$ (dashed lines) and 
$\epsilon$ $\!=$ $\!0.8$ (dotted lines). For comparison, the exact 
distributions that are observed in the limit $\epsilon$ $\!\to$ $\!0$ 
are also shown (solid lines).  }
\end{figure}

\begin{figure}
\centering\epsfig{figure=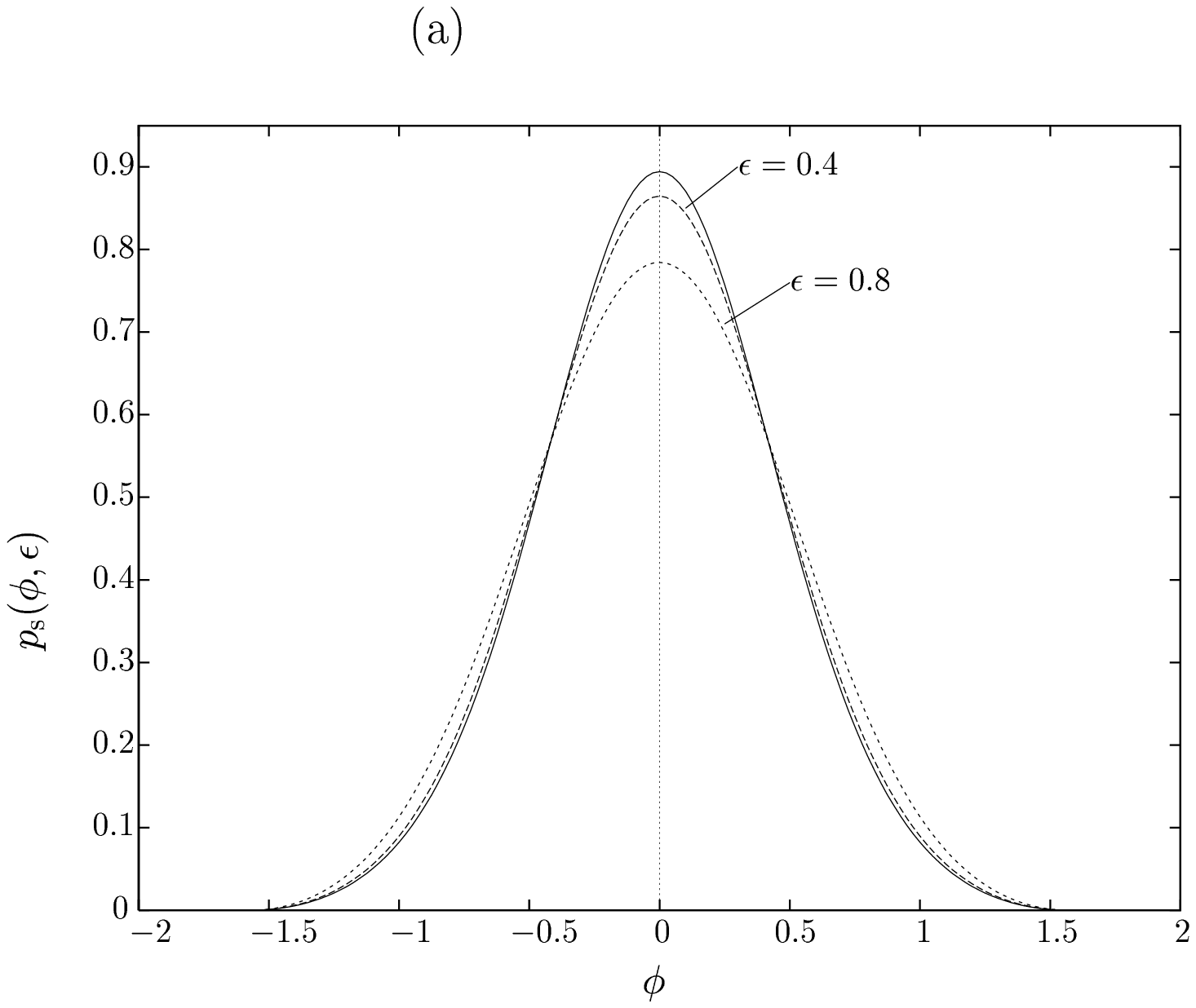,width=0.8\linewidth}
\centering\epsfig{figure=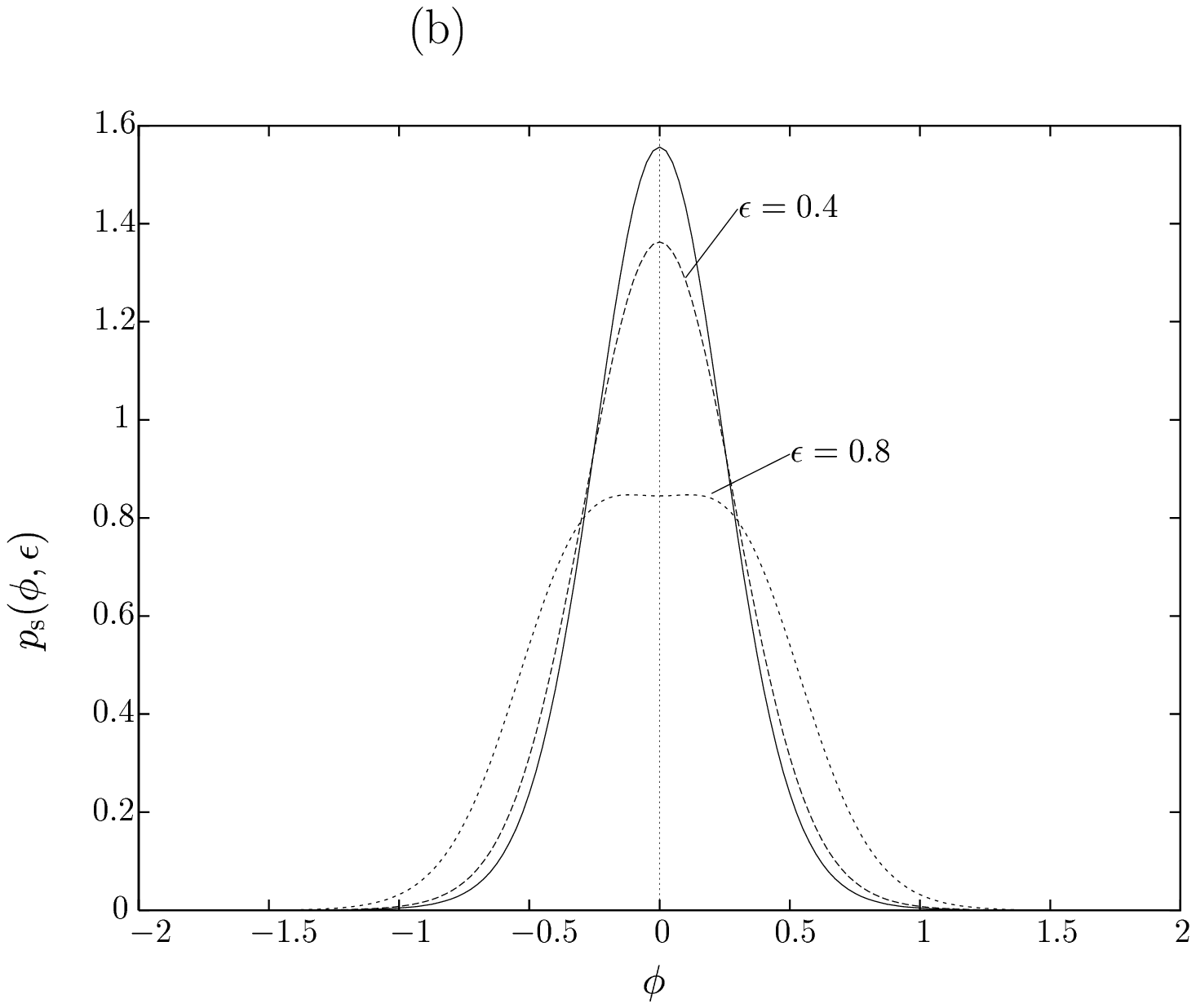,width=0.8\linewidth}
%\centering\epsfig{figure=pr13fig3.ps,width=0.8\linewidth}
%\centering\epsfig{figure=pr13fig4.ps,width=0.8\linewidth}
\caption{ \label{figr2} The coarse-grained sine phase state distributions of
a mode prepared in a coherent state $|\alpha\rangle$ with
$\langle\hat n\rangle$ $\!=$ $\!1$ $(\alpha$ $\!=$ $\!1)$ (a) and 
$\langle\hat n\rangle$ $\!=$ $\!2$ $(\alpha$ $\!=$ $\!\protect\sqrt{2})$ 
(b) are shown for $\epsilon$ $\!=$ $\!0.4$ (dashed lines) and 
$\epsilon$ $\!=$ $\!0.8$ (dotted lines). For comparison, the exact 
distributions that are observed in the limit $\epsilon$ $\!\to$ $\!0$ 
are also shown (solid lines). }
\end{figure}

The states $|\Phi,\psi,\epsilon\rangle$ 
can be used to define parametrized phase 
distributions of a radiation-field mode via their overlaps 
with the quantum state $\hat\varrho$ of the mode,
\begin{equation}
p(\Phi,\psi,\epsilon) = 
N^{-1}(\psi,\epsilon) \,
\langle\Phi,\psi,\epsilon|\,\hat{\varrho}\,|\Phi,\psi,\epsilon\rangle,
\label{1.10}
\end{equation}
where the normalization factor
\begin{equation}
N(\psi,\epsilon) = \int_{0}^{\pi} {\rm d}\Phi \,
\langle\Phi,\psi,\epsilon|\,\hat{\varrho}\,|\Phi,\psi,\epsilon\rangle
\label{1.12}
\end{equation}
has been introduced. 
Note that $p(\Phi,\psi,\epsilon)$ $\!\to$ $\!p(\Phi,\phi)$
$\!\equiv$ $\!\langle\Phi,\psi|\hat \varrho|\Phi,\psi\rangle$ 
when $\epsilon$ $\!\to$ $\!0$.
In particular, the coarse-grained cosine and sine phase state
distributions $p_{\rm c}(\phi,\epsilon)$ and $p_{\rm s}(\phi,\epsilon)$
are given by
\begin{equation}
p_{\rm c}(\phi,\epsilon)
= p(\Phi\!=\!\phi,\psi\!=\!0,\epsilon),
\label{1.13}
\end{equation}
\begin{equation}
p_{\rm s}(\phi,\epsilon)    
= p\!\left(\Phi\!=\!\textstyle\frac{1}{2}\pi\!-\!\phi,
\psi\!=\!\textstyle\frac{1}{2}\pi,\epsilon\right)
\label{1.14}
\end{equation}
[$p_{\rm c}(\phi)$ $\!\equiv$ $\!\langle\cos\phi|\hat\varrho|\cos\phi\rangle$
$\!=$ $\lim_{\epsilon\to 0}$ $\!p_{\rm c}(\phi,\epsilon)$,
$p_{\rm s}(\phi)$ $\!\equiv$ $\!\langle\sin\phi|\hat\varrho|\sin\phi\rangle$
$\!=$ $\lim_{\epsilon\to 0}$ $\!p_{\rm s}(\phi,\epsilon)$].
The smaller the value of $\epsilon$ becomes the better the
coarse-grained distributions approximate to the exact ones.
In Figs.~\ref{figr1}-- \ref{figr2} plots of 
$p_{\rm c}(\phi,\epsilon)$ and $p_{\rm s}(\phi,\epsilon)$ for various
values of $\epsilon$ are shown for a mode prepared in a coherent state.
We see that with increasing value of the mean number of photons the value 
of $\epsilon$ must be decreased in order to obtain the coarse-grained
distributions comparably close to the exact ones.
This is of course a reflection of the fact that with decreasing value
of $\epsilon$ the value of the (effective) cut-off photon number in the 
expansion (\ref{1.5a}) is increased.

\section{Relations to the phase-parametrized
field-strength distributions}
\label{sec3}
When we let $\epsilon$ $\!=$ $\!0$, 
then Eq.~(\ref{1.5a}) reduces, after multiplication by $\epsilon^{-1/2}$,
to the expansion in the photon-number basis of the exact 
states $|\Phi,\psi\rangle$.
All the photon-number states are seen to
contribute to the exact states 
with comparable weight, which 
prevents one from directly sampling the exact distributions from the 
difference-count statistics recorded in balanced homodyning.   
The advantage of the coarse-grained distributions
is that they can be directly sampled from the recorded data.  
Since for sufficiently small values of the coarse-graining parameter the 
measured distributions approximate to the exact distributions with
desired accuracy, the method enables us to asymptotically measure the 
exact distributions.

As already mentioned, the difference-count statistics recorded in balanced 
homodyning represents the statistics of a scaled field strength of the
signal mode,
\begin{equation}
\hat{F}(\varphi) = |F|(\hat{a}e^{-i\varphi}+\hat{a}^\dagger e^{i\varphi}),
\label{3.0}
\end{equation}
where the phase $\varphi$ is determined by the chosen phase parameters of 
the apparatus (see, e.g., \cite{Vogel3}). The desired sampling functions
can be therefore obtained from the integral relations between the
coarse-grained phase distributions 
$p(\Phi,\psi,\epsilon)$
and the phase-parametrized field-strength distributions of the
signal field,
\begin{equation}
p({\cal F},\varphi)=
\langle {\cal F},\varphi |\,\hat{\varrho}\,| {\cal F},\varphi\rangle,
\label{3.1}
\end{equation}
$|{\cal F},\varphi\rangle$  being the eigenvectors of
$\hat{F}(\varphi)$ (for details, see \cite{Vogel3,Schubert2}).
For this purpose we note that the single-mode density operator can be
expanded as \cite{Cahill1,KVogel1,Ariano3,Leonhardt1}
\begin{equation}
\hat{\varrho} =  \int_0^\pi {d}\varphi \int_{-\infty}^\infty {d} {\cal F}
\,p({\cal F},\varphi;s) \, \hat{K}({\cal F},\varphi;-s),
\label{3.2}
\end{equation}
where the smeared field-strength distributions $p({\cal F},\varphi;s)$,
$s$ $\!=$ $\!1\!-\!\eta^{-1}$, have been introduced which are measured
in non perfect detection, i.e., when the detection efficiency $\eta$
is less than unity (see, e.g., \cite{Vogel3}). 
The operator integral kernel $\hat{K}({\cal F},\varphi;-s)$ in 
Eq.~(\ref{3.2}) is given by
\begin{equation}
\hat{K}({\cal F},\varphi;-s)
= \,\frac{|F|^2}{\pi} \int_{-\infty}^\infty {d} y \, |y|\,
\exp\!\left\{iy\left[\hat{F}(\varphi)\!-\!{\cal F}\right]
-{\textstyle\frac{1}{2}}sy^2|F|^2\right\}\!.
\label{3.3}
\end{equation}
Using Eq.~(\ref{3.2}), from Eq.~(\ref{1.10})
we easily find  that $p(\Phi,\psi,\epsilon)$
can be related to $p({\cal F},\varphi;s)$ as
\begin{equation}
p(\Phi,\psi,\epsilon) = N^{-1}(\psi,\epsilon)
\int_{0}^{\pi}d\varphi\int_{-\infty}^{\infty} {d}{\cal F}\,
p({\cal F},\varphi;s)\, K_\epsilon(\Phi,\psi,{\cal F},\varphi;s),
\label{3.4}
\end{equation}
where
\begin{equation}
K_\epsilon(\Phi,\psi,{\cal F},\varphi;s)=
      \langle\Phi,\psi,\epsilon|\,\hat{K}({\cal F},\varphi;-s)\,|
      \Phi,\psi,\epsilon\rangle.
\label{3.6a}
\end{equation}

Equation~(\ref{3.4}) can be regarded as the basic equation for direct 
sampling of the phase distributions $p(\Phi,\psi,\epsilon)$ 
from the difference-count statistics in balanced homodyning,
where the integral kernel play the role of the sampling function.
In order to calculate them, we substitute in Eq.~(\ref{3.6a}) for 
$|\Phi,\psi,\epsilon\rangle$ the expansion Eq.~(\ref{1.5a}), and we derive
\begin{eqnarray}
\lefteqn{
K_\epsilon(\Phi,\psi,{\cal F},\varphi;s) = \frac{2\epsilon}{\pi}
\sum_{n=0}^{\infty}\sum_{m=0}^{\infty}
\Big\{
f_{nm}(x;s)\,\exp[i(n\!-\!m)(\varphi\!-\!\psi)]
}
\nonumber \\ && \hspace{2ex}\times \,
\sin[(n\!+\!1)\Phi]\sin[(m\!+\!1)\Phi]
%\nonumber \\ && \hspace{12ex} \times \,
\,{\rm sinc}[(n\!+\!1)\epsilon/2]\,{\rm sinc}[(m\!+\!1)\epsilon/2]
\Big\},
\label{3.7}
\end{eqnarray}
where the function
$f_{nm}(x;s)$, $x$ $\!=$ $\!{\cal F}/(\sqrt{2}|F|)$, is closely
related to the sampling function 
\begin{equation}
      \langle n|\hat{K}({\cal F},\varphi;-s)|m\rangle=
      f_{nm}(x;s)\exp[i(n-m)\varphi]
\label{3.9}
\end{equation}
for measuring the signal-mode density matrix
in the photon-number basis \cite{Ariano3,Leonhardt1}. From 
inspection of Eqs.~(\ref{3.3}) and (\ref{3.7}) we see that
the symmetry relations
\begin{equation}
K_\epsilon(\Phi,\psi,{\cal F},\varphi\!+\!\pi;s)
= K_\epsilon(\Phi,\psi,-{\cal F},\varphi;s),
\label{3.9b}
\end{equation}
\begin{equation}
K_\epsilon(\Phi,\psi,{\cal F},\varphi\!+\!\pi;s)
= K_\epsilon(\Phi\!-\!\pi,\psi,{\cal F},\varphi;s),
\label{3.9c}
\end{equation}
\begin{equation}
K_\epsilon(-\Phi,\psi,{\cal F},\varphi;s)
= K_\epsilon(\Phi,\psi,{\cal F},\varphi;s),
\label{3.9d}
\end{equation}
\begin{equation}
K_\epsilon(\Phi,-\psi,{\cal F},-\varphi;s)
= K_\epsilon(\Phi,\psi,{\cal F},\varphi;s),
\label{3.9da}
\end{equation}
\begin{equation}
K_\epsilon(\Phi,\psi,{\cal F},\varphi;s)
= K_\epsilon(\Phi,0,{\cal F},\varphi\!-\!\psi;s)
\label{3.9e}
\end{equation}
are valid. Hence knowing the function 
\begin{equation}
K_\epsilon(\Phi,{\cal F},\varphi;s)
\equiv K_\epsilon(\Phi,\psi\!=\!0,{\cal F},\varphi;s), 
\label{3.9f}
\end{equation}
with $\Phi,\varphi \in (\pi/2)$-intervals,
the function $\!K_\epsilon(\Phi,\psi,{\cal F},\varphi;s)$ is known 
for all values of $\Phi$, $\psi$, and $\varphi$. In particular, 
the functions $K_\epsilon^{\rm c}(\phi,{\cal F},\varphi;s)$
and $K_\epsilon^{\rm s}(\phi,{\cal F},\varphi;s)$, respectively,
that are required in order to relate the the cosine and sine 
phase states distributions $p_{\rm c}(\phi,\epsilon)$ and
$p_{\rm s}(\phi,\epsilon)$ to the field-strength distributions 
are given by
\begin{equation}
K_\epsilon^{\rm c}(\phi,{\cal F},\varphi;s)=
     K_\epsilon(\Phi\!=\!\phi,{\cal F},\varphi;s),
\label{3.5}
\end{equation}
\begin{equation}
K_\epsilon^{\rm s}(\phi,{\cal F},\varphi;s)=
     K_\epsilon(\Phi\!=\!\phi\!-\!\textstyle\frac{1}{2}\pi,{\cal F},
\varphi\!-\!\frac{1}{2}\pi;s).
\label{3.6}
\end{equation}

\section{Direct sampling of the Susskind-Glogower phase distributions}
\label{sec4}
The function $f_{nm}(x;s)$ in the series expansion (\ref{3.7}) of the 
sampling function $K_\epsilon(\Phi,{\cal F},\varphi;s)$, Eq.~(\ref{3.9f}), 
has been studied in a number of papers and different algorithms 
for numerical calculations have been discussed (see \cite{Richter1,Leonhardt2} 
and references therein). For the sake of transparency let us restrict 
attention to perfect detection ($\eta$ $\!=$ $\!1$ and hence 
$s$ $\!=$ $\!1$ $\!-$ $\!\eta^{-1}$ $\!=$ $\!0$). In this case, 
$f_{nm}(x)$ $\!\equiv$ $\!f_{nm}(x,s\!=\!0)$ can be written as
\begin{equation}
f_{nm}(x) = \frac{\rm d}{{\rm d} x} 
\left[ \psi_n(x) \phi_m(x) \right] \quad {\rm if} \quad m \geq n
\label{3.9a}
\end{equation}
[$f_{nm}(x)$ $\!=$ $\!f_{mn}(x)$ if $m$ $\!<$ $n$],
where $\psi_n(x)$ and $\phi_m(x)$, respectively, are the regular
(normalizable) and irregular (unnormalizable) solutions of the 
energy eigenvalue equation of the harmonic oscillator 
for the $n$th eigenvalue \cite{Richter1,Leonhardt2}.
The asymptotic behaviour of $f_{nm}(x)$ for large values of $n$ and $m$
can be found using the semiclassical (WKB) approximation. For the 
argument $x$ within the classical allowed region
$|x|$ $\!<$ $\!a_n$ $\!\equiv$ $\!(2n+1)^{\frac{1}{2}}$ the function 
$f_{nm}(x)$ ($m$ $\!\geq$ $\!n$) becomes \cite{Leonhardt2}
\begin{eqnarray}
\lefteqn{
f_{nm}(x) \sim \frac{2}{\pi} \, (p_{n}p_m)^{-\frac{1}{2}}
\left[p_m\cos\!\left(S_n\!+\!\textstyle\frac{1}{4}\pi\right)
\cos\!\left(S_m\!+\!\textstyle\frac{1}{4}\pi\right)
\right.
}
\nonumber \\ && \hspace{16ex}
\left.
- \, p_n\sin\!\left(S_n\!+\!\textstyle\frac{1}{4}\pi\right)
\sin\!\left(S_m\!+\!\textstyle\frac{1}{4}\pi\right)
\right]\!,
\label{3.10}
\end{eqnarray}
where
\begin{equation}
\label{3.10a}
p_n(x) = \left(2n+1-x^2\right)^{\frac{1}{2}}
\end{equation}
denotes the classical momentum and
\begin{equation}
\label{3.10b}
S_n(x) = \int_{a_n}^x {\rm d}x' \, p_n(x')
\end{equation}
is the time-independent part of the  classical action.

\begin{figure}
\noindent
\begin{minipage}[b]{0.5\linewidth}
\centering\epsfig{figure=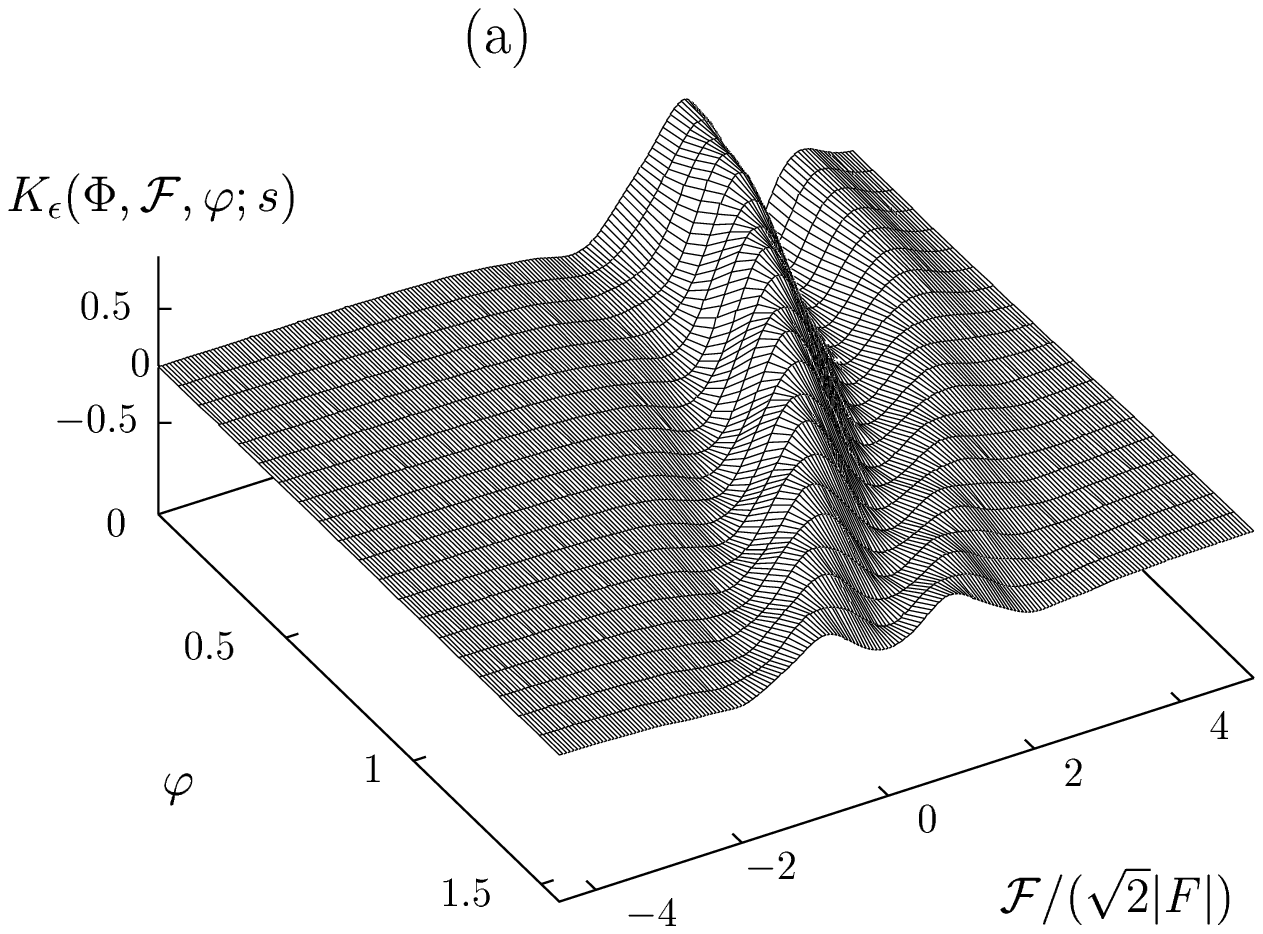,width=\linewidth}
%\centering\epsfig{figure=pr13fig5.ps,width=\linewidth}
\end{minipage}\hfill
\begin{minipage}[b]{0.5\linewidth}
\centering\epsfig{figure=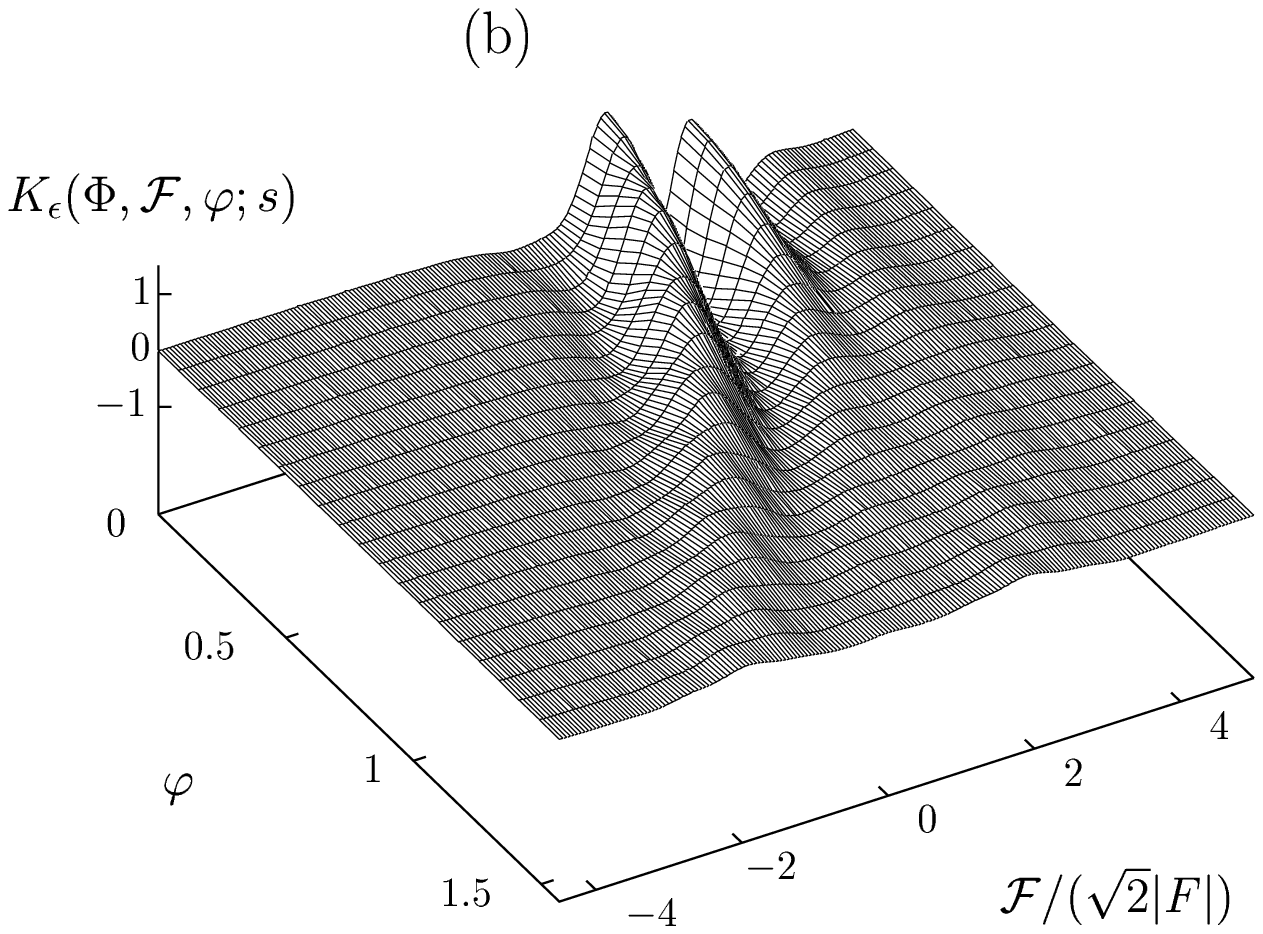,width=\linewidth}
%\centering\epsfig{figure=pr13fig6.ps,width=\linewidth}
\end{minipage}

~

\vspace{3cm}

~

\begin{minipage}[b]{0.5\linewidth}
\centering\epsfig{figure=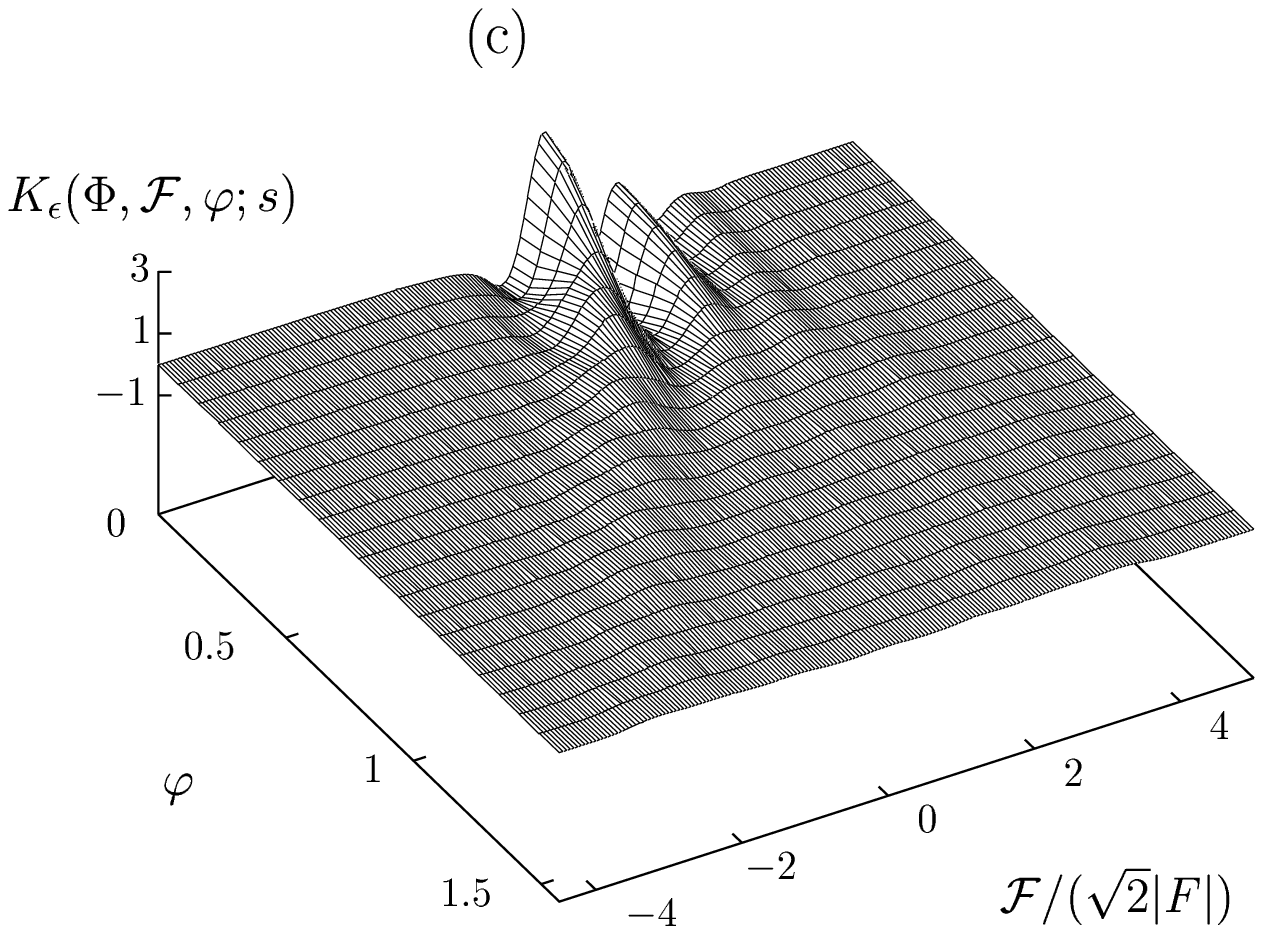,width=\linewidth}
%\centering\epsfig{figure=pr13fig7.ps,width=\linewidth}
\end{minipage}\hfill
\begin{minipage}[b]{0.5\linewidth}
\centering\epsfig{figure=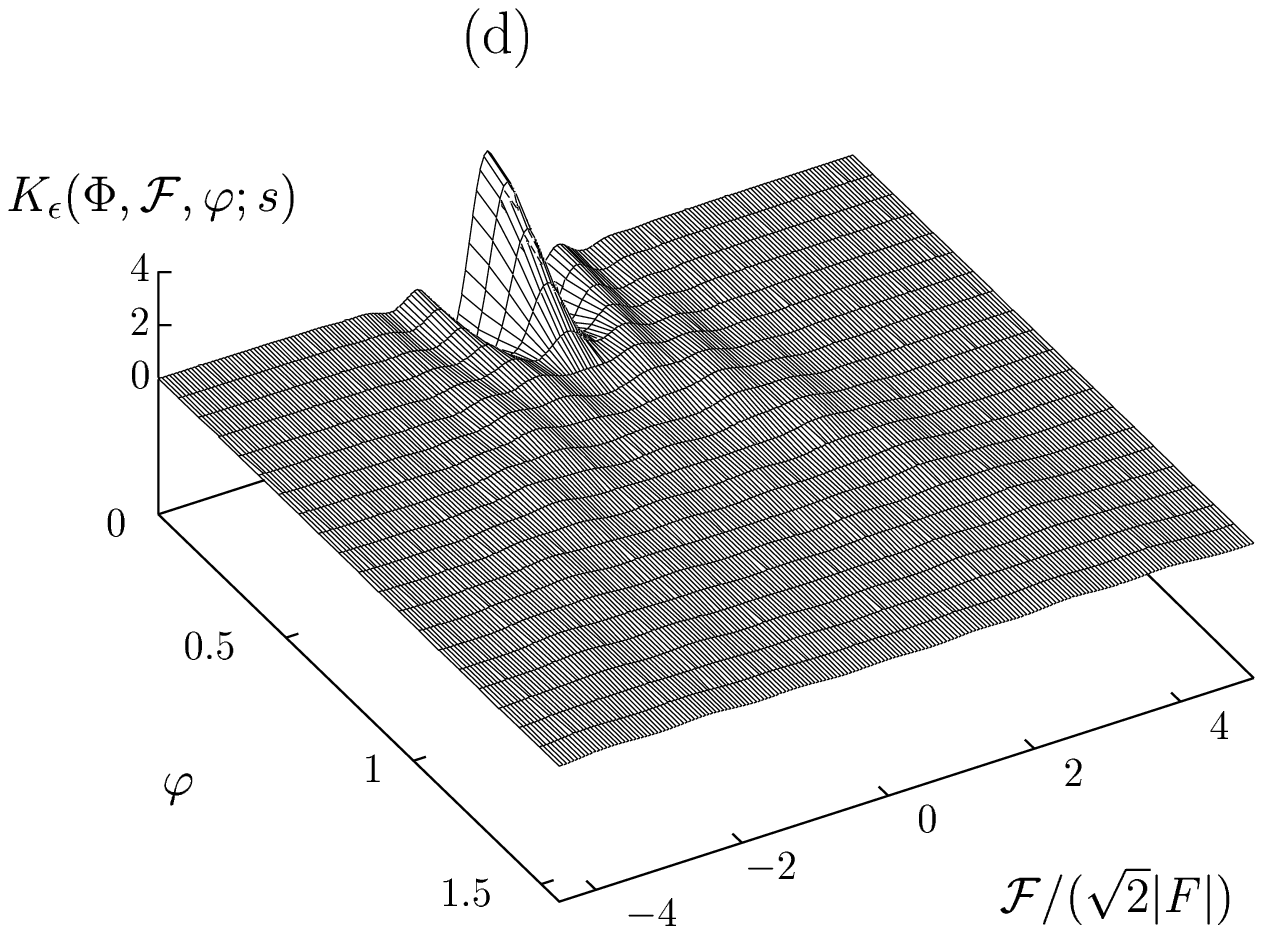,width=\linewidth}
%\centering\epsfig{figure=pr13fig8.ps,width=\linewidth}
\end{minipage}
\caption{ \label{figr3} The dependence on ${\cal F}$ and $\varphi$ of the 
sampling function $K_\epsilon(\Phi,{\cal F},\varphi;s)$
is shown for various values of $\Phi$, the values of $s$ and $\epsilon$
being $s$ $\!=$ $\!0$ (perfect detection) and $\epsilon$
$\!=$ $\!0.4$; %. \protect\\
$\Phi$ $\!=$ $\!\frac{1}{8}\pi$ (a),
%\protect\\
$\Phi$ $\!=$ $\!\frac{1}{4}\pi$ (b),
%\protect\\
$\Phi$ $\!=$ $\!\frac{3}{8}\pi$ (c),
%\protect\\
$\Phi$ $\!=$ $\!\frac{1}{2}\pi$ (d). }
\newpage
\end{figure}

For proving the convergence of the series expansion of
$K(\Phi,{\cal F},\varphi;s\!=\!0)$, it is sufficient to 
substitute in Eq.~(\ref{3.7}) for $f_{nm}(x;s\!=\!0)$ the semiclassical 
expression (\ref{3.10}). For $x/a_n$ $\!\to$ $\!0$
the functions $p_n(x)$ and $S_n(x)$ in Eq.~(\ref{3.10}) behave
like $n^{1/2}$ and $-(n\!+\!\frac{1}{2})\pi/2$, respectively, so that
$(p_m(x)/p_n(x))^\frac{1}{2}$ $\!\sim$ $\!(m/n)^\frac{1}{4}$
and $\cos[S_n(x)\!+\!\pi/4]$ $\!\sim$ $\cos(n\pi/2)$,  
$\sin[S_n(x)\!+\!\pi/4]$ $\!\sim$ $-\sin(n\pi/2)$. Hence for
any $\epsilon$ $\!>$ $\!0$ the series expansion is expected
to exist, because of the factor $(nm)^{-1}$ that arises from the
sinc-functions. Results of numerical calculations of 
$K(\Phi,{\cal F},\varphi;s\!=\!0)$ are shown in Fig.~\ref{figr3}.
It should be noted that $K(\Phi,{\cal F},\varphi;s)$ separately depends
on the two phases $\Phi$ and $\varphi$, whereas the sampling function for 
the London phase state distribution only depends on the difference phase
\cite{Dakna1}.

In order to demonstrate the feasibility of direct sampling of the
SG cosine and sine phase state distributions from the recorded
difference count statistics in balanced homodyning, we have performed
computer simulations of measurements of the phase-parametrized
field-strength distributions $p({\cal F},\varphi)$ on an equidistant
grid of points $\{{\cal F}_i,\varphi_j\}$. We have
assumed that in the experiments the signal mode to be detected 
is prepared in a squeezed vacuum state
\begin{equation}
|\Psi\rangle = \exp\!\left\{
-\textstyle\frac{1}{2} \left[
\xi (\hat a ^\dagger)^2 - \xi^\ast \hat a^2
\right]
\right\} |0\rangle.
\label{4.1}
\end{equation}

\begin{figure}
\centering\epsfig{figure=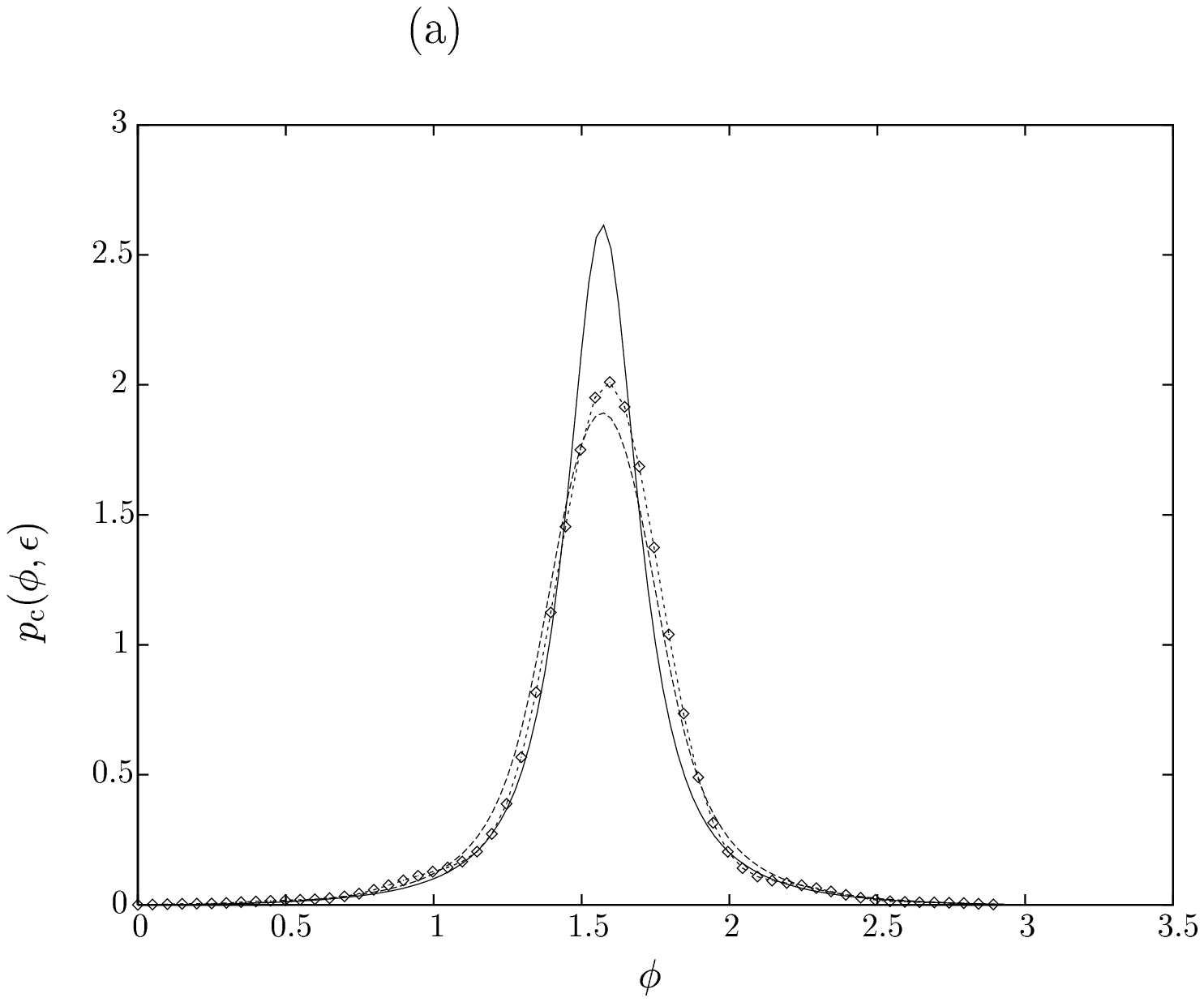,width=0.8\linewidth}
\centering\epsfig{figure=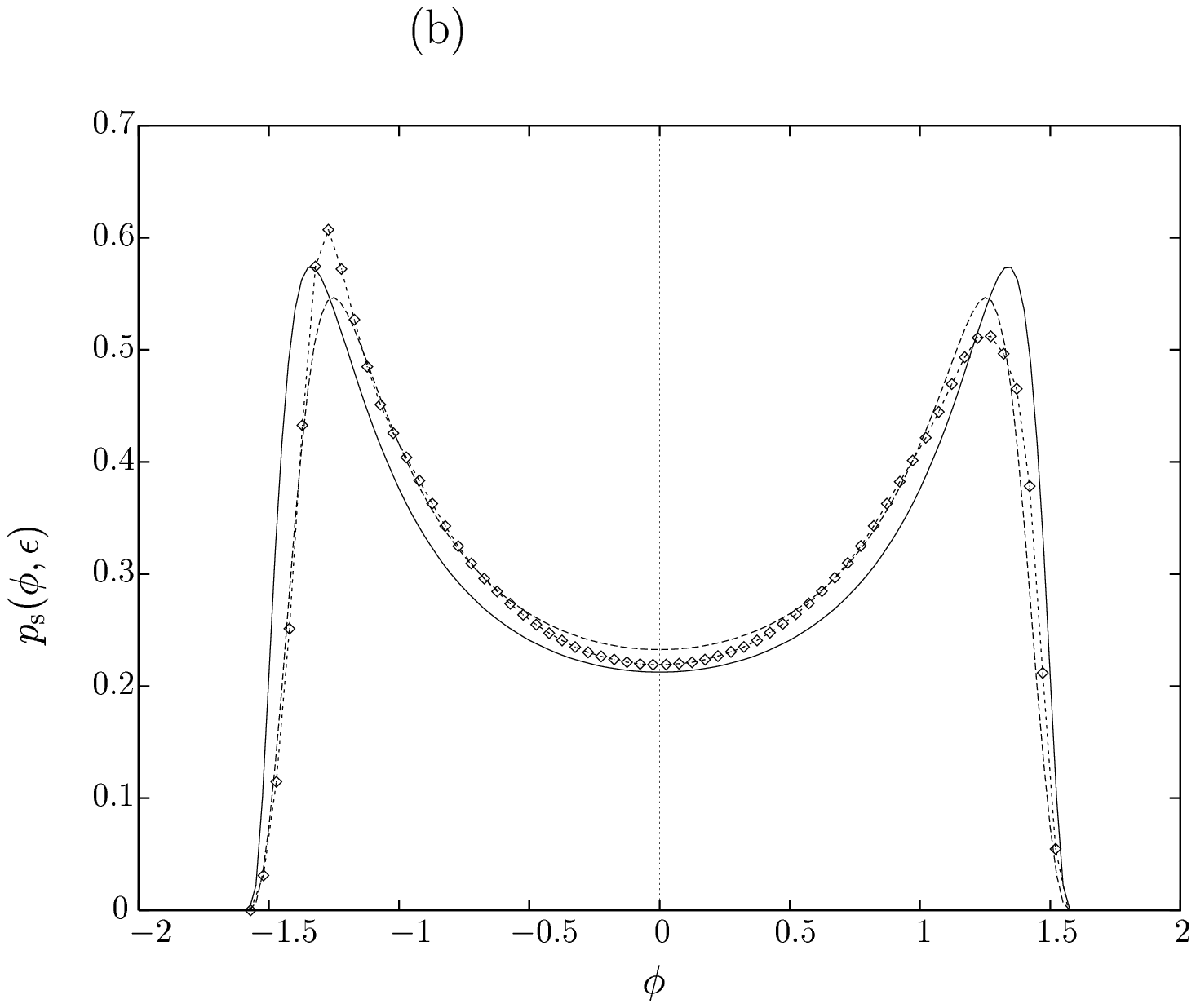,width=0.8\linewidth}
%\centering\epsfig{figure=pr13fig9.ps,width=0.8\linewidth}
%\centering\epsfig{figure=pr13fig10.ps,width=0.8\linewidth}
\caption{ \label{figr4} The measured
Susskind--Glogower cosine (a) and sine (b) phase state distributions 
(double dotted lines) of a signal mode prepared in a squeezed vacuum state, 
Eq.~(\protect\ref{4.1}), with mean photon number $\langle\hat n\rangle$ 
$\!=$ $\!1$ ($\xi$ $\!=$ $\!0.88$) are compared with the calculated distributions
(dashed lines), the values of $s$ and $\epsilon$ being 
$s$ $\!=$ $\!0$ (perfect detection) and $\epsilon$ $\!=$ $\!0.4$. 
The solid lines represent the exact distributions that 
are observed in the limit $\epsilon$ $\!\to$ $\!0$. In the computer
simulation of measurements $10^4\times 30$ events ($30$ phase values)
are assumed to be recorded.  }
\end{figure}

In particular, we have assumed that $10^4\times 30$ events 
are recorded. From the results shown in Fig.~\ref{figr4} we see that 
the sampled distributions are in good agreement with the theoretical 
predictions. Their deviations from the exact distributions obviously 
result from the middle observational level chosen. Nevertheless, the 
accuracy is seen to be sufficient in order to detect the typical features
of the exact distributions. From Eqs.~(\ref{1.4d}) [together with 
Eqs.(\ref{1.4e}) and (\ref{1.4f})] and (\ref{4.1}) we easily see that
the cosine and sine phase state distributions of the ordinary vacuum
($\xi$ $\!=$ $\!0$) are given by $|\langle 0|\cos\phi\rangle|^2$ $\!=$ 
$\!(2/\pi)\sin^2\phi$ and $|\langle 0|\sin\phi\rangle|^2$ $\!=$ 
$\!(2\pi)^{-1}\cos^2\phi$, respectively. In this case the cosine phase 
state distribution has a broad maximum at $\phi$ $\!=$ $\!\pi/2$, whereas 
the sine phase state distribution has a broad maximum at $\phi$ $\!=$ $\!0$.
When the value of $\xi$ is increased ($\xi$ $\!>$ $\!0$), the 
``vacuum noise circle'' centred at the origin of co-ordinates in the 
(complex) $\alpha$ phase space is squeezed to an ellipse with the small 
semi-axis parallel to the real axis \cite{Loudon1}. Hence, the 
cosine phase distribution is expected to become more sharply peaked at
$\phi$ $\!=$ $\!\pi/2$. Accordingly, the sine phase state distribution
is expected to show a double-peak structure, with the maxima close to
$\phi$ $\!=$ $\!\pm \pi/2$ and the minimum at $\phi$ $\!=$ $\!0$.

\section{Summary and conclusions} 
\label{sec5}
In the paper we
have studied the problem of measuring the SG cosine and sine
phase state distributions in balanced homodyne detection. We have 
shown that they can be directly sampled, in a coarse-graining
approximation, from the recorded difference-count statistics with
sufficiently well accuracy. The accuracy is determined by the interval 
of field strengths used for ``probing'' the phase statistics of the 
signal-mode under consideration. With decreasing coarse-graining parameter, 
this interval is increased and the accuracy is increased as well.

It is worth noting that the SG cosine and sine phase state 
distributions can be regarded as special cases of $\psi$-parametrized
phase state distributions. The latter are based on $\psi$-parametrized
phase states that for $\psi$ $\!=$ $\!0$ and $\psi$ $\!=$ $\!\frac{1}{2}
\pi$ reduce to the cosine and sine phase states, respectively.
Accordingly, the sampling functions for the cosine and sine phase
state distributions can be obtained by specifying the
sampling function for the $\psi$-parametrized phase distributions.
Their integral relation to the field-strength distributions
reveal that the sampling function exhibits a number of symmetry
properties that can advantageously be used in calculations.

We have calculated the
sampling function using an expansion in terms of the matrix elements
of the corresponding operator integral kernel in the Fock basis.
These matrix elements, which are the sampling functions required for 
measuring the signal-mode density matrix in the Fock basis, can be 
calculated applying the algorithm developed in \cite{Leonhardt2}. In 
this way, the sampling function can be obtained for any (non-zero) 
coarse-graining parameter.

In order to give an example, we have performed computer simulations of
measurements of the cosine and sine phase state distributions
at a middle observational level (i.e., for a not extremely small
coarse-graining parameter and a realistic grid of sampling
points). Assuming that the signal mode is prepared in a squeezed
vacuum state, the measured cosine and sine phase state distributions have 
been found to show the typical features of the exact distributions. Apart 
from the maxima and minima indicating the regions of nonclassical noise 
reduction and enhancement, respectively, qualitatively different shapings 
of the two distributions are observed which also demonstrate the quantum 
character of the state of the signal mode. It should be pointed
out that the different shapings reflect the fact that for quantum
states the (noncommuting) quantities cosine and sine cannot be related 
to a common phase. This effect is of course not observed in the London 
state phase distribution. Clearly, the introduction of a phase that 
cannot be related to a Hermitian operator is the price that is to be 
paid for a common phase. 
 
\section{Acknowledgements}
This work was supported by the Deutsche Forschungsgemeinschaft.
We would like to thank T. Opatrn\'{y} and U. Leonhardt for discussions.

\newpage
\newpage
\newpage
\end{document}